\documentstyle[prl,aps,graphicx]{revtex}

\begin{document}
\draft

\twocolumn
\narrowtext

\title{Polaron Effects on Superexchange Interaction: \\
Isotope Shifts of $T_N$, $T_C$, and $T^*$ in Layered Copper Oxides}

\author{M. V. Eremin, I. M. Eremin, I. A. Larionov, and A. V. Terzi}

\address{Physics Department, Kazan State University, Kazan 420008, Russia \\
e-mail: Mikhail.Eremin@ksu.ru}

\maketitle
\date{February 20, 2002}

\begin{abstract}
A compact expression has been obtained for the superexchange coupling of
magnetic ions via intermediate anions with regard to polaron effects at both
magnetic ions and intermediate anions. This expression is used to analyze
the main features of the behavior of isotope shifts for temperatures of
three types in layered cuprates: the Neel temperatures ($T_N$), critical
temperatures of transitions to a superconducting state ($T_C$), and
characteristic temperatures of the pseudogap in the normal state ($T^*$).
\end{abstract}



Elucidating the nature of unusual isotope effect in copper-oxygen
superconductors is one of the most important problems on the way to
ascertaining the mechanism of the pairing of charge carriers in these
compounds. It is known that the observation of an isotope shift of the
superconducting transition temperature ($T_C$) was of crucial importance in
ascertaining the phonon mechanism of pairing in conventional
(low-temperature) superconductors. The fact that an isotope effect exists in
high-temperature superconductors built up of copper–oxygen planes has long
been beyond question; however, the relative smallness of the coefficient
(for example, $\alpha \sim 0.056$ for YBa$_2$Cu$_4$O$_8$ upon replacing
$^{16}$O with $^{18}$O, instead of the standard value 0.5) and the specific
features of its behavior in other compounds upon changing the number of
holes in copper-oxygen planes do not fit the Bardeen-Cooper-Schrieffer (BCS)
scenario. At the same time, the majority of authors of articles related to
the isotope effect (see the recent review [1]) correctly point out that,
nevertheless, phonon modes, in some mysterious way, affect the
superconducting transition temperature. In this context, we believe that the
facts of observing an isotope effect for the characteristic temperature of
the pseudogap state of underdoped cuprates (so-called pseudogap onset
temperature $T^*$) gain great importance. Thus, according to [2], the
isotope exponent upon replacing $^{16}$O with $^{18}$O corresponding to
$T^*$ is = 0.061. It was natural that this fact suggested a common origin of
$\alpha_{T_C}$ and $\alpha_{T^*}$ [2].

We believe that another and even more important similarity in the dependence
of the order parameters of the superconducting and pseudogap phases on the
$d$ - type wave vector (that is, $\cos q_x - \cos q_y$) was explained in its
time under the assumption that the transitions to both these phases are
associated with short-range potentials [3]. It seems that superexchange
interaction, screened Coulomb repulsion, and interaction of holes mediated
by optical phonons are the most significant of these. The isotope shift of
$T_C$ and $T^*$ due to interaction mediated by optical phonons was discussed
in a few works (see, for example [4,5]), and that for $T^*$ was considered
in [6]. Below, we will focus our attention on polaron corrections to the
superexchange coupling of copper spins ($J$) and will demonstrate that a
number of features in the behavior of the isotope shift of $T_C$ upon
changing the number of holes in the CuO$_2$ plane can be quite reasonably
explained even within the framework of the purely superexchange mechanism of
pairing. First, we emphasize the following important fact. Within the
scenario [3], the superconducting transition temperature $T_C \sim 2J - G$,
whereas the characteristic temperature of the pseudogap phase $T^* \sim J+G$
[3,7]. Here, $G$ is the parameter of the screened Coulomb interaction of
holes on the nearest copper sites, which partially includes the correction
due to interaction mediated by optical phonon modes. If it is granted that
the isotope shift is associated with the phonon renormalization of $G$, the
shifts of $T_C$ and $T^*$ would be of opposite sign, which is in
contradiction with the experimental results [2]. In this connection, it is
believed that the scenario of phonon renormalization of the parameter $J$ is
more reasonable to suggest as the source of positive isotope shifts of $T_C$
and $T^*$.

The renormalization of $J$ within the Hubbard model was investigated in
detail by Kugel' and Khomskii [8]. It is evident from the above estimates
$T_C \sim 2 J - G $ and $ T^* \sim J + G$ that the small increase in $J$
proportional to the phonon frequency found in this work gives the correct
sign of isotope shifts but does not provide the required magnitude of the
shifts. This can be most simply demonstrated with the example of the isotope
shift of the Neel temperature in related high-$T_C$ compounds. As was
already indicated in [1], the Kugel' and Khomskii's correction gives a
correct sign of the isotope shift of the Neel temperature ($T_N$) in
La$_2$CuO$_4$ upon replacing $^{16}$O with $^{18}$O; however, even the most
overrated estimates give a value that is six times lower than the
experimental one.

We believe that the main reason for the quantitative disagreement between
the theory [8] and the experiment [1] is in the fact that the Hubbard model
is not suitable for the compounds of our interest. As was already stressed
in [9], that the energies of electron transfer from oxygen to a magnetic ion
($\Delta_c$) in the majority of copper oxides are smaller than the energy of
electron transfer from copper to copper ($\Delta_a$). However, the Hubbard
model gives correct estimates only when $\Delta_a \geq \Delta_c$.

The Hamiltonian in the form
\begin{eqnarray}
\label{eq1}
\hat{H} = \sum{\varepsilon_{a} a^{+}_{\sigma} a_{\sigma}} +
\sum{\varepsilon_{c} c_{\sigma}^{+} c_{\sigma}} +
\sum{U_{a} n^{a}_{\uparrow} n^{a}_{\downarrow}} +  \nonumber \\
+ \sum {U_{c}n^{c}_{\uparrow}n^{c}_{\downarrow}} +
\sum{t_{ac}\left({a_{\sigma}^{+}c_{\sigma}+c_{\sigma}^{+}a_{\sigma}}\right)}.
\end{eqnarray}
is best suited to the description of superexchange interaction with taking
into account explicitly the cascade hopping of electrons over oxygen. As
applied to high-$T_C$ superconductors, it is rather frequently named the
Emery Hamiltonian. Here, $t_{ac}$ is the hopping integral between
neighboring copper and oxygen sites, and $U_a$ and $U_c$ are the parameters
of electron Coulomb repulsion. We will estimate the corrections to the
superexchange parameter $J$ due to polaron effects at copper (a) and oxygen
(c) sites in the fashion of [8], supplementing Eq. (1) with the
electron-phonon coupling operator
\begin{equation}
\label{eq2}
\hat{H}_{ep} = \sum \limits_{a,b,c} {\mathrm g}_{i}n_{i}
\left({p_{q}+p^{+}_{-q}} \right).
\end{equation}
Here, $p_q$ and $p_{-q}^{+}$ are phonon annihilation and creation operators,
and ${\mathrm g}_i$ is the coupling parameter connected with the polaron
stabilization energy ($E_i$) at site $i$ by the equation $E_i = {\mathrm
g}_i^2 / \hbar \omega_i $, where $\omega_i$ are local vibrational
frequencies.

In the physical context, this calculation corresponds to a simplified
Holstein model when the migrating charge locally interact with breathing
modes, forming electron–vibrational states with dispersionless optical
phonons. In this connection, it is pertinent to note that conduction in the
compounds under consideration is just of the polaron type and is
accomplished mainly via oxygen ion sites [10].

This calculation is most simply performed by the method of canonical
transformations. The matrix of the unitary transformation of the initial
Hamiltonian is found by excluding the odd terms with respect to hopping
integrals with an accuracy of to sixth-order perturbation theory. The
calculation, whose mathematical details will be given in a more detailed
article, gives the following result:
\begin{eqnarray}
\label{eq3}
J = J_{0} {\Bigg ( }1+ \frac{3\hbar}{\left(\Delta_{ca} \right)^{2}} {\Big [
} E_{a} \omega_{a} \coth \left(\frac{\hbar \omega_{a}}{2k_{B}T} \right)+
\nonumber \\
E_{c} \omega_{c} \coth\left(\frac{\hbar \omega_{c}}{2k_{B}T}
\right) {\Big ]} {\Bigg ) },
\end{eqnarray}
where $\Delta_{ac} = \varepsilon_a - \varepsilon_c + U_a - U_c$ has the
meaning of the energy of transfer from oxygen to copper, and the corrections
proportional to $E_a \hbar \omega_a / \Delta_{ac} U_a^2 $ and $E_a \hbar
\omega_a / U_a^3 $ are not given because of their smallness for the compounds under
consideration. $J_0$ is the parameter of superexchange interaction of copper
spins via the intermediate oxygen atom in the absence of phonons [9,11].
Note that the appearance of temperature factors in our equation is generally
characteristic for the problems on transitions in transition metal compounds
with the participation of quasilocal vibrations [12]. At the same time, one
should keep in mind that polaron effects break down at $T \sim \omega $, and
the concepts used here become inapplicable.

The results of our calculation are given in the figure. The system of
integral equations for the mean field parameters corresponding to the
transition to the pseudogap phase was solved self-consistently. We identify
the pseudogap phase with the phase of sliding charge-density waves. This
system was written in detail in [3] and is not given here.

For the description of the superexchange coupling parameter upon replacing
some isotopes for other ones at $\hbar \omega / k_B T$, it is convenient to
introduce parameters $\gamma_{Cu}$ and $\gamma_O$ by writing
\begin{equation}
\label{eq4}
J = J_{0}\left[1+\gamma_{Cu}\left(\frac{\Delta M_{Cu}}{M_{Cu}}\right)
+\gamma_{O}\left(\frac{\Delta M_{O}}{M_{O}}\right)\right].
\end{equation}
It follows from Eq. (3) that, upon replacing $^{16}$O with $^{18}$O,
\begin{equation}
\label{eq5}
\gamma_{O} \approx - \frac{3}{2}
\left(\frac{E_{a}\hbar\omega_{a}}{\Delta^{2}_{Cu-O}}\right),
\end{equation}
whereas, upon replacing $^{63}$Cu with $^{65}$Cu,
\begin{equation}
\label{eq6}
\gamma_{Cu} \approx
-\frac{3}{2}\left(\frac{E_{c}\hbar\omega_{c}}{\Delta^{2}_{Cu-O}}\right).
\end{equation}
Substituting here (in electronvolts) $\Delta_{Cu-O}$ = 1.5 [13] and standard
values $\hbar \omega_a$ = 0.05, $E_a$ = 0.4 [10,14], and using the
relationship $\Delta T_N/T_N \approx \Delta J / J $ characteristic for
layered cuprates (see [1]), we find that the Neel temperature should
decrease by 0.2$\%$ upon replacing $^{16}$O with $^{18}$O in
YBa$_2$Cu$_3$O$_{6.383}$. According to measurements in La$_2$CuO$_4$ [15],
the shift $\simeq 0.6 \%$. If, however, it is assumed, following [1], that
$E_a$ = 1.2 eV, our estimate will coincide with the experimental value. We
hope that this explanation of the isotope shift of $T_N$ will stimulate
further experimental investigations of this important problem. Our estimated
value $\gamma_O \approx$ $-$ 0.014 is overrated. The value $\gamma_O
\approx$ $-$ 0.01 is better suited to comparison of the calculated with experiment,
see the figure.

It is pertinent to note that, generally speaking, there is another
possibility of changing $J$ given by Eq. (4). This possibility is associated
with the change in the distance between copper ions upon replacing some
isotopes with other ones. It is known that the superexchange parameters very
strongly depend on the distance between the interacting ions. This mechanism
explains well the increase in $T_C$ under the action of an external pressure
on a high-$T_C$ crystal [16,17]. The question naturally arises in this case
of what occurs with the lattice parameters upon changing some isotopes for
other ones. Recent precise measurements in YBa$_2$Cu$_4$O$_8$ crystal showed
[18] that the lattice parameters $a$, $b$, and $c$ in the case of $^{16}$O
equal (in ${\mathrm \AA }$) 3.8411(1), 3.8717(1), and 27.2372(8),
respectively, whereas these are equal to 3.8408(1), 3.8718(1), and
27.2366(8), respectively, for $^{18}$O, that is, these parameters are
somewhat smaller in the latter case. The positive isotope shift of the
nuclear quadrupole resonance frequency of plane copper nuclei [18] is
another important experimental fact, which indicates that interatomic
copper–oxygen distances are smaller in the case of $^{18}$O. Based on these
data, one may only conclude that the change of interatomic distances upon
replacing some isotopes with other ones must lead to negative shifts of
$T_C$ and $T^*$ and will, probably, be relatively small. From the
theoretical point of view, this fact seems quite understandable, because
changes in interatomic distances upon replacing some isotopes with other
ones are due to the lattice anharmonicity, and its effect is naturally of
less importance than the effect of harmonic vibrations.

\begin{figure}  [tbp]
\centering\includegraphics[height=0.7\linewidth,width=\linewidth]{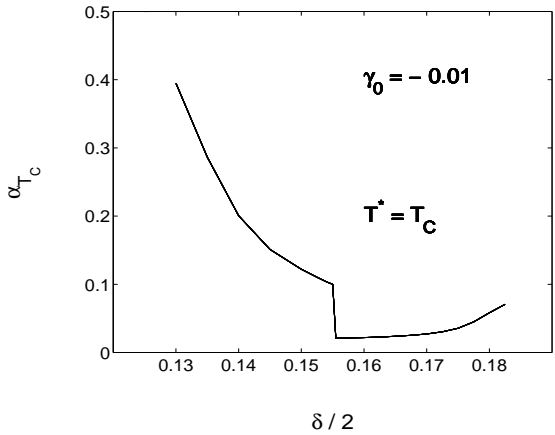}
\caption{The calculated exponent of the isotope shift of $T_C$ upon replacing $^{16}$O
with $^{18}$O as a function of the number of holes per one copper site.}
\label{Figure}
\end{figure}

In the figure, the values of the coefficient $\alpha_{T_C} = - d
\ln (T_C) / d \ln (M)$ for the replacement of $^{16}$O with $^{18}$O
are plotted as abscissas, and the numbers of holes per one copper site are
plotted as ordinates. The symbols in the curve correspond to the points at
which the system of self-consistent equations from [3] was solved. Because
only the order of magnitude is known for the polaron energies $E_a$ and
$E_c$, the parameter $\gamma_O$ was normalized in such a way that
$\alpha_{T_C}$ was equal to 0.1 at the optimal level of doping. The
calculated behavior of $\alpha_{T_C}$ is not symmetric with respect to the
point of optimal doping. The physical nature of this asymmetry is associated
with strong competition between $d$-SC and $id$-CDW phases in the underdoped
state. On the left of this point, the isotope shift exponent increases with
decreasing number of holes, approaching 0.5, whereas the value of
$\alpha_{T_C}$ remains virtually constant as well as it does at $\delta$/2
greater than 0.16 (so-called overdoped regime). This kind of asymmetry (but
without a step!) in the behavior of $\alpha_{T_C}$ as a function of the
number of holes was found recently in measurements [19]. The authors of this
work mentioned already that, if only conventional interaction via the phonon
field were responsible for the isotope effect and the unusual drop in the
optimal doping region were related to a peak in the density of states, the
curve would be approximately symmetric with respect to the point of optimal
doping. At $\delta/2$ larger than 0.16, the value of $\alpha_{T_C}$ would
strongly increase; however, this was not found [19].

As to the isotope shift of $T_C$ upon replacing the copper $^{63}$Cu isotope
with $^{65}$Cu or $^{66}$Cu, the fact noted in [20] that the ratio
$\alpha_{T_C}$(Cu)/$\alpha_{T_C}$(O) $\approx$ 0.75 $\pm$ 0.1 does not
depend on the type of the compound and on the doping level is naturally
explained based on Eq. (3). The stabilization energy of a small-radius
polaron (hole) at a copper site is higher than that at an oxygen site. The
nearest environment of a hole at an oxygen site comprises positive copper
ions, whereas the nearest environment of copper comprises negative oxygen
ions. This fact is the reason for the difference between $\gamma_O$ and
$\gamma_{Cu}$.

The value of $\alpha_{T^*}$ calculated in this work in the region 0.1 $<$
$\delta/2$ $<$ 0.16 turned out to be approximately constant: $\alpha_{T^*}
\approx$ 0.01. This is smaller than the value estimated in experiments
(0.061) [2]; therefore, the effect of interaction via optical phonons on
$\alpha_{T^*}$ discussed in [2] cannot be excluded. This is also
corroborated by a number of experimental points in Fig. 2 from [19] on the
right of the point of optimal doping. We hope to describe this problem in a
more detailed work.

Thus, the renormalization of the superexchange interaction of copper spins
due to polaron effects noted in this work explains the main regularities of
the isotope shift of the superconducting transition temperature in layered
cuprates both in the order of magnitude and in the sign and the character of
the dependence on the number of holes. The starting equation for the
renormalization of the superexchange parameter was verified using the
isotope shifts of the Neel temperature of the parent compounds as an
example. Our calculations were based on the scenario of competition between
the superconducting phase and the charge-density-wave phase. Agreement
between the calculations and experiment confirms this scenario. At the same
time, our calculation predicts a rather sharp jump of the isotope exponent
$\alpha_{T_C}$ on passing through the point of optimal doping. This effect
is relatively small; however, we believe that the experimental observation
of this effect would be of principal importance.

This work was supported by the Russian Program Superconductivity, project
No. 98014-2, CRDF Rec 007 and INTAS project No. YSF 2001/2-45, and partially
supported by Swiss National Scientific Foundation, project No. 7SUPJ062258.

\end{document}